\documentclass[a4paper]{article}

\usepackage[utf8]{inputenc}
\usepackage{multicol}
\usepackage{caption}
\usepackage{float}

\usepackage[brazilian]{babel}
\usepackage{graphicx}
\usepackage{hyperref}
\hypersetup{
	setpagesize  = false,
	colorlinks   = true,    % Colours links instead of ugly boxes
	urlcolor     = blue,    % Colour for external hyperlinks
	linkcolor    = black,   % Colour of internal links
	citecolor    = black    % Colour of citations
}

\usepackage{authblk}
\usepackage{amsmath}

\newcommand{\keywordsenglishname}{Keywords}

\renewenvironment{abstract}{%
	\begin{center}
		\begin{minipage}{14cm}
			{\textbf{\abstractname:}}
		}{
		\end{minipage}
	\end{center}
}
\newenvironment{abstractinenglish}{
	\def\abstractname{\abstractinenglishname}
	\begin{abstract}
	}{
	\end{abstract}
}
\newenvironment{keywords}{
	\def\abstractname{\emph{\keywordsportugues}}
	\begin{abstract}
	}{
	\end{abstract}
}
\newenvironment{keywordsenglish}{
	\def\abstractname{\emph{\keywordsenglishname}}
	\begin{abstract}
	}{
	\end{abstract}
}

\title {\vspace{-2.5cm} Motor elétrico -- SimuFísica\textsuperscript{\textregistered}: um aplicativo para o ensino de eletromagnetismo\\[1ex] \large Electric Motor -- SimuFísica\textsuperscript{\textregistered}: an application for teaching electromagnetism}
\author{Marco P. M. de Souza\thanks{Endereço de correspondência: marcopolo@unir.br} $^1$ }
\author{Sidnei P. Oliveira$^{1,2}$}
\author{Valdenice L. Luiz$^{1}$}
\affil{$^1$Departamento de Física, Universidade Federal de Rondônia, Ji-Paraná, RO, Brasil.}
\affil{$^2$Secretaria de Estado da Educação -- SEDUC, RO, Brasil.}
\date{}

\usepackage{fancyhdr}
\begin{document}	
	
	\maketitle	
	
	\begin{abstract}
		Apresentamos neste trabalho o simulador Motor elétrico, um aplicativo da plataforma SimuFísica\textsuperscript{\textregistered} voltado para uso em sala de aula. Descrevemos brevemente as tecnologias por trás do aplicativo, as equações que regem o seu funcionamento, alguns estudos mostrando a dinâmica do motor elétrico e, por fim, exemplos de abordagem em sala de aula.
	\end{abstract}
	
	
	\vspace{6pt}
	
	\begin{abstractinenglish}
		In this work, we present the Electric Motor simulator, an application on the SimuFísica\textsuperscript{\textregistered} platform intended for use in the classroom. We briefly describe the technologies behind the application, the equations that govern its operation, some studies showing the dynamics of the electric motor and, finally, examples of approach in the classroom.
	\end{abstractinenglish}
	
	\begin{keywordsenglish}
		Electric motor, computer simulation, Application, Simulator, Electromagnetim.
	\end{keywordsenglish}

	\vspace{1cm}
	
	\begin{multicols}{2}

	\section{Introdução}
	
	A temática do motor elétrico pode ser um ponto de partida eficaz para ensinar eletromagnetismo devido aos vários tópicos desse assunto relacionados ao seu funcionamento. Sendo uma aplicação prática dos princípios do eletromagnetismo, os alunos podem ver diretamente como as leis do eletromagnetismo se manifestam no mundo real, tornando o assunto mais concreto e relevante. O estudo do motor elétrico requer a compreensão e integração de vários conceitos de eletromagnetismo, como corrente elétrica, campo magnético, força magnética, espiras e bobinas, e também da mecânica, como rotação, torque e aceleração angular.
	
	Experimentos envolvendo a aplicação do motor elétrico no ensino de Física tem sido um tema recorrente na literatura, aparecendo em vários artigos \cite{Schubert, Hudha, Diniz, Pires, Monteiro} e dissertações \cite{Silva, Euzebio}. Por outro lado, simulações computacionais de um motor elétrico com fins didáticos são bastante escassas. Isso nos motivou a desenvolver e apresentar o aplicativo Motor elétrico, disponível na plataforma SimuFísica\footnote{\url{https://simufisica.com/}}, que abordamos na próximas seções.

	\begin{figure*}[ht]
		\centering
		\includegraphics[width=0.6\linewidth]{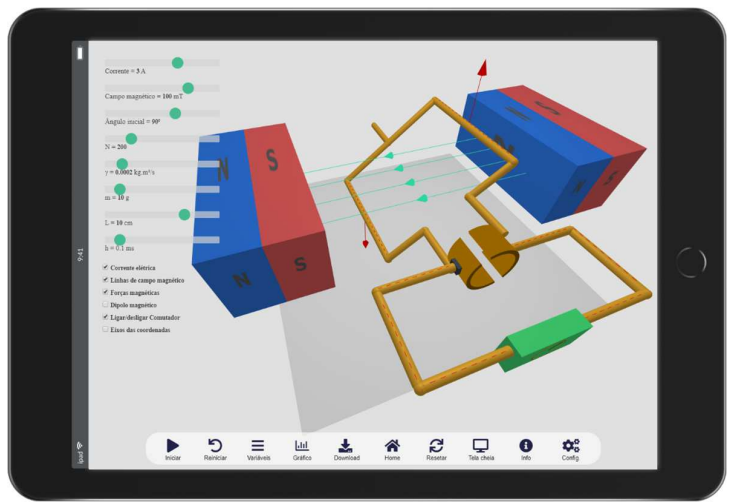}
		\caption{Versão 1.6.6 do aplicativo Motor elétrico aberto em um iPad 6. As flechas vermelhas e verdes representam, respectivamente, as forças magnéticas que agem em cada espira e as linhas de campo magnético entre os ímãs. Disponível através do link \url{https://simufisica.com/simulacoes/motor-eletrico/}.}
		\label{fig1}
	\end{figure*}

	\section{A plataforma SimuFísica\textsuperscript{\textregistered}}
	
	O SimuFísica\textsuperscript{\textregistered} é uma coleção de aplicativos simuladores voltados para o ensino de Física em nível médio e superior. Com sua natureza multilíngue e multiplataforma, o SimuFísica\textsuperscript{\textregistered} possui simulações que abrangem assuntos como o consumo de energia elétrica, o gás ideal e a propagação da função de onda da mecânica quântica, por exemplo. Esses simuladores podem ser acessados online ou instalados em computadores, tablets e smartphones de diversos sistemas operacionais.
	
	Tendo como público-alvo estudantes e professores, o uso do SimuFísica\textsuperscript{\textregistered} em sala de aula, aliado a um bom planejamento, pode trazer vantagens significativas para o ensino de Física. A plataforma oferece a oportunidade de visualizar fenômenos físicos de forma interativa, facilitando a compreensão de conceitos abstratos. Além disso, os alunos podem realizar ``experimentos'' virtuais, explorando diferentes parâmetros e observando as consequências rapidamente. Isso promove o aprendizado ativo, estimula o pensamento crítico e o desenvolvimento de habilidades de resolução de problemas. O \textit{feedback} imediato proporcionado pelo SimuFísica\textsuperscript{\textregistered} permite aos alunos corrigir erros e aprimorar o aprendizado. A plataforma também oferece acesso flexível, sendo disponível online e na versão de aplicativo para dispositivos móveis e desktop, permitindo que alunos e professores acessem os simuladores em qualquer lugar e a qualquer momento. Ao incorporar o SimuFísica\textsuperscript{\textregistered} aos planos de aula, os professores podem enriquecer a instrução em sala de aula, complementar a teoria com exemplos práticos e ajudar os alunos a visualizar as aplicações da Física no mundo real.

	\section{O aplicativo Motor elétrico}
	
	Trata-se da simulação de um motor elétrico de corrente contínua composto por uma bobina com $N$ espiras retangulares na presença de um campo magnético gerado por um par de ímãs (Fig. \ref{fig1}). O usuário pode configurar a intensidade da corrente elétrica $i$, a intensidade do campo magnético $\vec{B}$, o ângulo inicial $\theta_0$ que o plano das espiras fazem com $\vec{B}$, o número de espiras, a constante de amortecimento $\gamma$ que age no rotação da espira, a massa $m$ da bobina, a largura $L$ das espiras e o parâmetro $h$, que regula o passo de integração numérica do sistema de equações diferenciais ordinárias (EDOs) que regem a dinâmica do sistema. O usuário tem a opção de visualizar tanto algumas variáveis numericamente, como frequência média de oscilação da bobina e seu fluxo magnético, bem como seus gráficos de evolução temporal. É possível, também, observar uma representação da corrente elétrica, do vetor força magnética total $\vec{F}_B$ agindo em dois dos lados das espiras e das linhas de campo magnético que atravessam a bobina.

	A versão online do \textit{app} Motor elétrico, assim como todos os outros da plataforma SimuFísica\textsuperscript{\textregistered}, tem seu código-fonte baseado nas três das principais tecnologias da web: HTML5 (conteúdo, isto é, \textit{tags} HTML), CSS (\textit{layout}) e Javascript (dinâmica, basicamente). O Javascript é responsável pela solução das EDO's a partir dos parâmetros e da condição inicial definidos pelo usuário, pela renderização em tempo real da simulação na tela através da manipulação da \textit{tag} Canvas e pela adequação do aplicativo aos dispositivos de diversos tamanhos, como desktops, tablets e smartphones. A simulação em 3D do \textit{app} é realizada através da biblioteca Three.js.
	
	As versões nos idiomas inglês e espanhol são obtidas por uma tradução que é feita, principalmente, de forma automática e localmente (no computador de desenvolvimento) através do framework Node.js e	
	da API (\textit{application programming interface} -- interface de programação de aplicações) Google translate, disponível de forma gratuita no gerenciador de pacotes npm (\textit{Node Package Manager}). Já as versões instaláveis obtidas via download na própria plataforma online ou nas lojas de aplicativos são desenvolvidas através das IDEs (\textit{Integrated Development Environment} -- Ambiente de Desenvolvimento Integrado) Android Studio (loja de aplicativo Play Store, do Google) e Xcode (App Store, Apple), ou com o \textit{framework} Electron.js (Microsoft Store $\rightarrow$ Windows,  e Snap Store $\rightarrow$ Linux). A Fig. \ref{fig2} mostra uma ilustração simplificada das principais tecnologias envolvidas no desenvolvimento das várias versões.

	{
		\centering
		\includegraphics[width=0.95\linewidth]{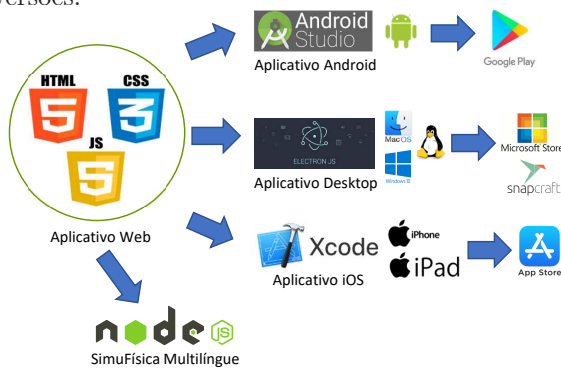}
		\captionof{figure}{Fluxograma simplificado de desenvolvimento do \textit{app}.}
		\label{fig2}
	}

	\section{A dinâmica}
	
	A dinâmica do motor elétrico é dada pela segunda lei de Newton para rotações com dissipação proporcional ao módulo da velocidade angular:
	
	\begin{equation}
		\tau = I\ddot{\theta} + \gamma\dot{\theta},
	\end{equation}
	
	\noindent onde $\tau$ é o valor absoluto do torque agindo na bobina, que possui distribuição uniforme de massa e cujo momento de inércia, calculado através da sua definição, é dado por
	
	\begin{equation}
		I = \dfrac{mW^2}{L+W} \left( \dfrac{L}{4} + \dfrac{W}{12}  \right),
	\end{equation}

	\noindent onde $L$ (definido pelo usuário) e $W = 10$ cm (valor fixo) são as dimensões das espiras. Sabendo-se que a magnitude do torque das forças magnéticas é dada por
	
	\begin{equation}
		\tau = F_BL\sin{\theta},
	\end{equation}
	
	\noindent chegamos no seguinte sistema de EDOs:
	
	\begin{subequations}
		\begin{align}
			\dot{\theta} &= \omega\\
			\dot{\omega} &= \dfrac{LF_B}{I}\sin{\theta} - \dfrac{\gamma}{I}\omega,
		\end{align}
	\end{subequations}
	
	\noindent que são integradas numericamente pelo método de Runge-Kutta de quarta ordem com passo temporal $h$.
	
	O campo magnético entre os ímãs é considerado uniforme e a intensidade da força magnética na bobina, composta por $N$ espiras (não mostradas na simulação), é dada por
	
	\begin{equation}
		F_B = iNLB.
	\end{equation}

	Na Fig. \ref{fig3} temos a evolução temporal do fluxo magnético total através das espiras. Consideramos uma bobina com $N = 200$ espiras, corrente elétrica $i = 3$ A, campo magnético $B = 100$ mT, ângulo inicial em relação ao sentido do campo magnético $\theta_0 = \pi/2$ rad, comprimento das espiras $L = 10$ cm, massa total $m = 10$ g e coeficiente de amortecimento $\gamma = 2\times 10^{-4}$ kg.m$^2$/s. Usamos um passo de integração de Runge-Kutta de $h = 0,1$ ms. Da figura podemos observar um fluxo magnético máximo com magnitude de 0,2 T.m$^2$. Em torno de $t = 10$ ms, e também em certos instantes posteriores, é observada uma alteração brusca no fluxo magnético, que vem da atuação do comutador. Esse dispositivo permite a troca do sentido da corrente nas espiras e, consequentemente, que o vetor dipolo magnético fique indefinidamente tentando se alinhar com o campo magnético dos ímãs, efeito que gera a rotação do motor.
	
	Podemos observar também que o período de oscilação do fluxo magnético apresentado na Fig. \ref{fig3} vai diminuindo com o passar do tempo. Isso depende, naturalmente, da massa da bobina, do valor do coeficiente de amortecimento e do valor do torque, que faz com que a média da velocidade angular cresça até atingir o regime estacionário (Fig. \ref{fig4}).
	
	{
		\centering
		\includegraphics[width=0.95\linewidth]{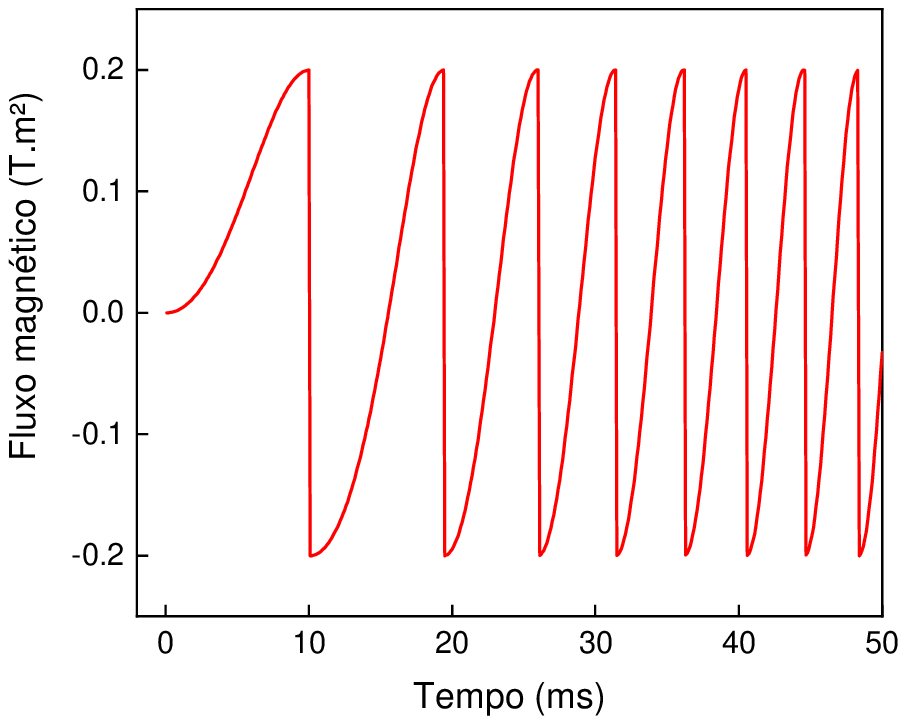}
		\captionof{figure}{Fluxo magnético nas espiras.}
		\label{fig3}
	}
	
	É interessante notar, também, a dinâmica do sistema sem a presença do comutador. Nesse caso, não temos, naturalmente, um motor elétrico. Fazendo $B = 100$ mT, $N = 100$, $\gamma = 10^{-3}$ kg.m$^2$/s, mantendo os outros parâmetros e condições iniciais e desabilitando o comutador no aplicativo, podemos observar na Fig. \ref{fig5} o torque sofrido pelas espiras e a tentativa destas de se alinharem com o campo magnético dos ímãs. É possível observar, pela Fig. \ref{fig5}(b), a presença de dois regimes conhecidos do oscilador harmônico simples \cite{Moyses}: amortecimento supercrítico (curva vermelha, onde a corrente vale $i = 0,1$ A) e amortecimento subcrítico (curva azul, onde a corrente vale $i = 10$ A).
	
	{
		\centering
		\includegraphics[width=0.95\linewidth]{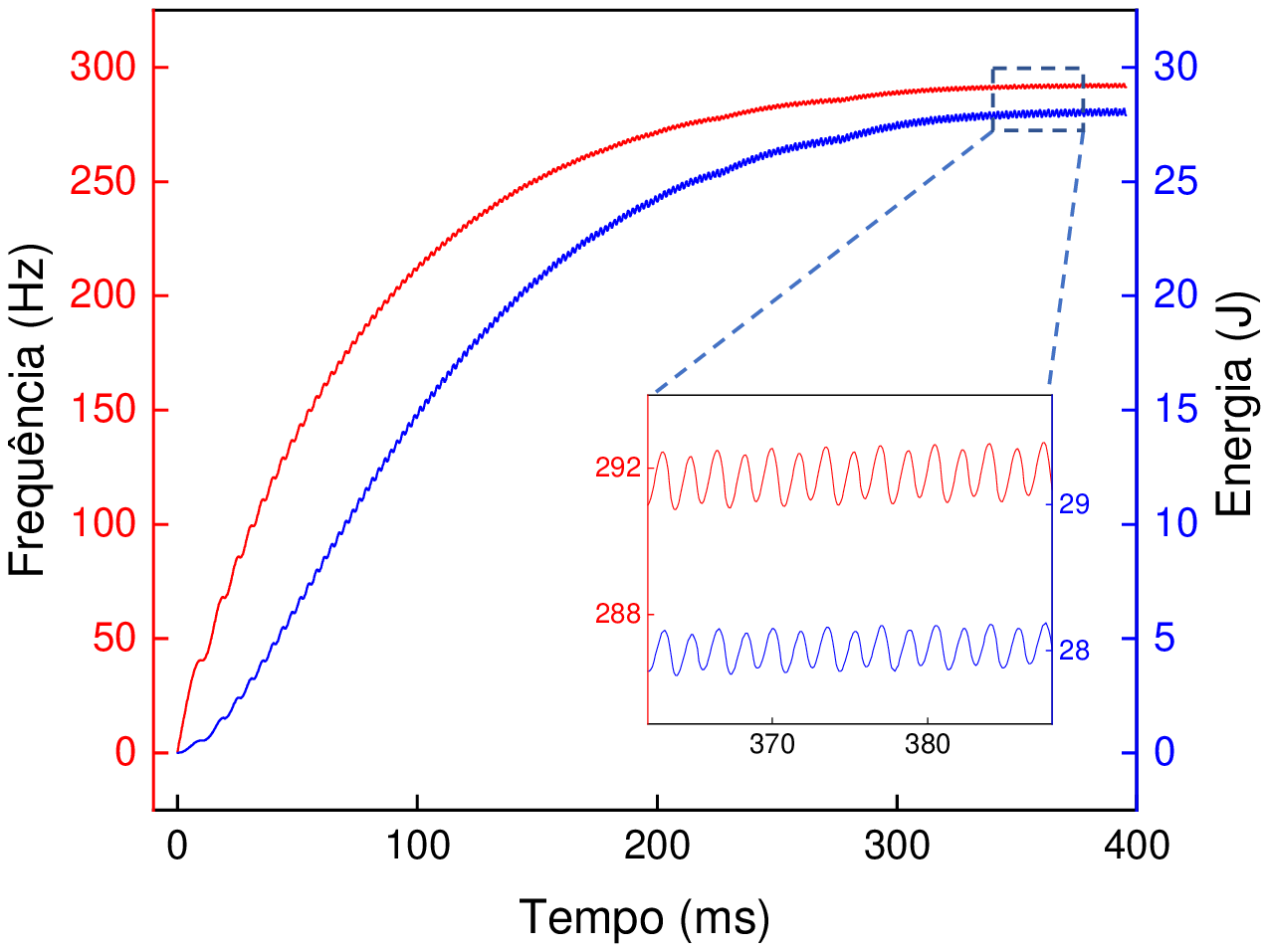}
		\captionof{figure}{Frequência de oscilação e energia cinética da bobina em função do tempo. \textit{Inset}: zoom da região retangular tracejada.}
		\label{fig4}
	}
	
	\section{Uso e abordagem em sala de aula}
	
	Por ser compatível com dispositivos móveis, é possível facilmente projetar, sem uso de cabos de vídeo, a simulação aberta no dispositivo em um datashow a partir do uso conjunto com uma TV box, como Apple TV (iOS e iPadOS) ou Roku Express (Android) \cite{Cristiane}, por exemplo, ou pelo próprio wifi do datashow, se disponível. %Estudos extra-classe
	
	Abaixo apresentamos três tópicos que podem ser abordados com o \textit{app}.

	\subsubsection*{Força magnética}
	
	A força magnética agindo em cada um dos dois lados das espiras que estão sempre perpendiculares à direção do campo magnético é dada por
	
	\begin{equation}
		\label{vetor-forca-magnetica}
		\vec{F}_B = i \vec{L} \times \vec{B}
	\end{equation}
	
	\noindent Conforme já informado, as três grandezas do lado direito da equação (\ref{vetor-forca-magnetica}) -- corrente elétrica, comprimento da espira e campo magnético -- podem ser variadas pelo usuário, de forma que o estudante pode facilmente visualizar o efeito das alterações dessas grandezas na força magnética que dois dos lados da espira experimentam.

	{
		\centering
		\includegraphics[width=0.95\linewidth]{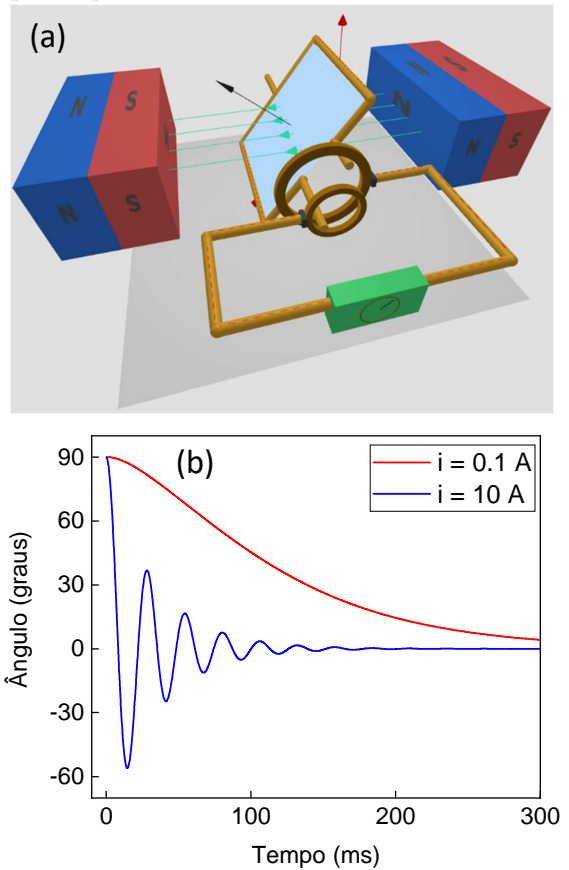}
		\captionof{figure}{(a) Simulação com o comutador desabilitado. A flecha preta representa o vetor dipolo magnético das bobinas. (b) Ângulo $\theta$ da bobina em função do tempo para duas correntes: 0,1 e 10 A.}
		\label{fig5}
	}

	\subsubsection*{Fluxo magnético}
	
	Como sabemos, o fluxo de campo magnético total $B$ sobre $N$ espiras de área $A = LW$ que mora em um dado plano do espaço é dado por
	
	\begin{equation}
		\label{fluxo-magnetico}
		\Phi_B = N B L W \cos{\theta},
	\end{equation}
	
	\noindent onde $\theta$ é o ângulo entre o vetor normal ao plano de espira e o vetor campo magnético, $L$ é o comprimento da espira e $W$ sua largura. Dessa vez, o usuário pode manipular quatro dos cinco parâmetros do lado direito da equação (\ref{fluxo-magnetico}).

	\subsubsection*{Momento de dipolo magnético e torque}
	
	Desabilitando o comutador no \textit{app}, é possível visualizar o toque $\tau$ que força o alinhamento de dipolos magnéticos $\vec{\mu}$ na direção de um campo magnético uniforme $\vec{B}$ (Fig. \ref{fig5}):
	
	\begin{equation}
		\label{torque}
		\vec{\tau} = \vec{\mu} \times \vec{B}
	\end{equation}
	
	\noindent Aqui, o usuário pode observar como um campo magnético externo $\vec{B}$ influencia a dinâmica de uma espira com momento de dipolo magnético $\vec{\mu}$.

	\section{Considerações finais}
	
	O aplicativo Motor elétrico, hospedado na plataforma SimuFísica\textsuperscript{\textregistered}, pode ser uma ferramenta valiosa no ensino de eletromagnetismo em nível de ensino médio e superior. Por ser leve e com rápida inicialização, e por ser compatível com os superportáteis smartphones do mundo contemporâneo, esta aplicação pode ser usada com bastante praticidade no ambiente acadêmico, proporcionando aulas mais interessantes.

	\section*{Agradecimentos}
	
	Este trabalho foi financiado pelo Conselho Nacional de Desenvolvimento Científico e Tecnológico -- CNPq (processo 304017/2022-1).

	\end{multicols}

\end{document}